\documentclass[a4paper]{article}
\usepackage{url}
\usepackage[cm]{fullpage}
\begin{document}
\title{Functional Programming and Security}
\author{Yusuf Moosa Motara\\
Department of Computer Science,\\
Rhodes University\\
South Africa\\
\textsf{y.motara@ru.ac.za}}
\date{November 2011}
\maketitle

\begin{abstract}
This paper analyses the security contribution of typical functional-language features by examining them in the light of accepted information security principles.  Imperative and functional code are compared to illustrate various cases.  In conclusion, there may be an excellent case for the use of functional languages on the grounds of better security; however, empirical research should be done to validate this possibility.
\end{abstract}

\section*{Introduction}

The functional paradigm dates from the 1930's, when Alonzo Church created a system called lambda calculus for understanding computation in terms of functions\cite{HistoryOfLambdaCalculus}.  At the time, as is the case at present, the dominant way in which to understand computation was as a series of operations that changed (or \emph{mutated}) data.  Lambda calculus provided a theoretical basis for understanding computation as a series of variable transformations, called \emph{functions}, instead.  It is important to note that the transformation of a variable is not equivalent to the \emph{mutation} of data; at the end of the transformation, there is simply some new data, which is the transformation result.  In functional terms, we can say that the data is \emph{immutable}: once created, it cannot be mutated.  Furthermore, there is no logical reason for a variable to be restricted to being data only; therefore, a variable may be a function, and may be transformed and created and passed to a different function in the same way that a variable would be.  Both of these ideas --- that data is immutable, and that functions are data --- are important aspects of functional thought, and we can see most features of modern-day functional languages as extensions of these ideas.

The hypothesis of this paper is that the functional paradigm, and the feature-set common to most widely-known functional languages, leads to code that is by default more secure than comparable mainstream imperative code.  Tevis \& Hamilton suggested the functional paradigm briefly as a more secure alternative in \cite[pp.~4--5]{Tevis2004}, and Tevis followed this up with a deeper look in \cite{SecureFuncParadigm}, where specific reference was made to Haskell.  This paper expands on his work, going into more depth and citing typically-functional features, as well as making explicit comparisons to the imperative model.  The integration of some functional features into the imperative mainstream is discussed in \cite{FuncPurityJava}, where the authors attempt to create a version of Java that includes functionally-pure methods, and discuss the security benefits of this approach as they proceed.

\section*{Security principles}

To test the hypothesis, a good definition of ``security'' which is not specific to the imperative paradigm is necessary.  Whereas Tevis has used the chapters of the ``Secure Programming Cookbook in C and C++''\cite{SecureProgrammingCookbook} to derive security principles, I have selected the principles provided in the second edition of ``Writing Secure Code''\cite{WritingSecureCode}, by Howard \& LeBlanc.  The reason for this is that the former is specifically geared towards implementation (since it is a cookbook) in imperative languages, and provides implementation advice rather than \emph{principles} for security.  The latter, though it still contains a bias towards imperative languages, describes a theoretical framework that underlies the security recommendations that are provided.  Furthermore, \cite{WritingSecureCode} has been accepted by industry bodies such as OWASP\cite{OWASPSecureCodingPrinciples} as being an acceptable baseline for reasoning about code security, and is part of the recommended reading list for Microsoft's Security Development Lifecycle methodology\cite{SDLArticle}.

Suitable comparison languages are also needed, to make the case with reference to concrete examples.  I have chosen to use C\# 2\footnote{Note that these numbers refer to \emph{language} versions, not Common Language Runtime (CLR) or .Net versions!} and F\# 2 as imperative and functional languages respectively.  The functionality found in the latter is also found in most other functional languages, such as Haskell, Scala, and OCaml.  The former is widely-used, and implements a set of functionality that is common to other mainstream ``enterprise''-style languages, such as Java.  It is also a managed language, just as F\# is; this is necessary for a fair comparison, since comparing the security benefits of unmanaged code \emph{vs} managed code would almost certainly result in a ``win'' for the managed option\cite[535--6]{WritingSecureCode}.  Though C\# is currently at version 3, I have chosen to use version 2 because version 3 brings with it several functional-style changes\cite{CSharpIsFunctional}, such as support for lambda-expressions and expression-based programming in the form of LINQ\footnote{\textbf{L}anguage \textbf{In}tegrated \textbf{Q}uery}.  Other mainstream imperative languages, such as C++ or Java, do not have these features.  Therefore, to be generalisable to a wider set of readers who are familiar with imperative code in an alternative language, I have chosen to use an older version of C\#.  Another point that should be addressed is that F\# is not a \emph{pure} functional language, and one can use non-functional constructs from within it.  I will not use any such constructs in my examples.

For each of the principles that are covered, I have attempted to unpack the meaning of the principle \emph{as it applies to programming}.  Some principles have been left unchanged; however, others are different from a design or deployment perspective, but have the same root cause as far as programming is concerned\footnote{This conclusion seems to have been reached independently by \cite{OWASPSecureCodingPrinciples}, which leaves out certain design- or deployment-focused principles.}.  These latter principles have been rationalised as follows:

\begin{itemize}
\item \emph{Fix Security Issues Correctly} and \emph{Learn from Mistakes} both refer to the idea of understanding the root cause of a defect, and fixing that.  The former principle emphasises determining the root cause, and the latter emphasises process change to try and ensure that similar mistakes do not occur in future.

\item \emph{Backward Compatibility Will Always Give You Grief} and \emph{Employ Secure Defaults} both refer to changing, removing or deprecating features to provide a more secure product.

\item \emph{Never Depend on Security Through Obscurity Alone} is an advisory principle rather than a rule or guideline.  It is advisory since it takes no position on ``Security Through Obscurity'' itself, but merely says that obscurity (if employed) should not be the only security measure used.  The positive formulation of this would say that many defenses, possibly including security through obscurity, should be used.  This formulation is quite similar to the principle of \emph{Use Defense in Depth}, and the two principles will therefore be covered together.

\end{itemize}

\subsection*{Plan on Failure}
\label{sec:PlanOnFailure}

When a failure occurs in a particular method, the imperative programmer is typically faced with two possibilities\footnote{In C\#, there is also the option of using an \textsf{out} parameter (though this option is rarely exercised in industry); in C++, pointers or references are possibilities.  None of this functionality is available in Java.} to return a \textsf{null} value, or to throw an exception.  This is particularly true in cases where failures are not expected: for example, when a dependent system, which has historically always been available, goes down.  Throwing an exception is possible, but is typically a last-resort since it has serious performance implications for mainstream imperative languages like Java\cite[\S6.1]{JavaPerformanceTuning}\cite{ExperimentingWithExceptions} and C\#\cite{EfficientExceptions}.

The alternative is to return a \textsf{null} value, thus indicating that the desired result of the function call is unavailable.  However, this option has consequences of its own that make it difficult to recommend.  The first is that the \textsf{null} value \emph{implies} that an error may have occurred, but may also mean that the call succeeded and the correct value to be returned is \textsf{null}.  There is no way for the caller to determine which of these two semantics is the correct one, other than to make more method calls.  The second is that an imperative language does not typically force a caller to handle the \textsf{null} value immediately, and the value doesn't carry a stack trace with it.  An exception only occurs once the value is dereferenced, which could be in a completely different part of the program; this makes debugging very difficult.

In either case, the caller must remember to create an alternative line of logic to deal with an error in the called function.  Forgetting to do so leads to a violation of this security principle.

In functional languages, the difference between success and failure is made explicit by a \emph{discriminated union} (called \textsf{Option} in F\# and Scala, and \textsf{Maybe} in Haskell).  A discriminated union differentiates between a fixed number of cases, with each case optionally having a type associated with it.  For example, the definition of the \textsf{Option} discriminated union is:

\begin{verbatim}
let 'a Option =
| Some of 'a
| None
\end{verbatim}

It is typically used as shown in the following pseudocode:

\begin{verbatim}
let getData x =
    if dataIsOK x then
        let result = (* create result here *)
        Some result
    else
        None

match getData someInput with
| Some x -> // handle success case
| None -> // failure case
\end{verbatim}

The caller of the \textsf{getData} function is forced to acknowledge that the call may not be successful, in which case the \textsf{None} case will be invoked.  In the event that the call is successful, the \textsf{Some} class will be invoked instead.  Note that it is quite possible for the return value to be \textsf{Some null}; in this event, it is clear that the call has succeeded, and the correct value happens to be \textsf{null}.  However, in idiomatic functional code, there is no reason for the \textsf{null} value to appear at all.  Haskell, for example, has no conception of \textsf{null}\footnote{The closest that Haskell comes to \textsf{null} is by having the concept of \textsf{undefined} or ``bottom'', which differs significantly from \textsf{null} in terms of behaviour; for example, any evaluation of \textsf{undefined}, whether in the context of comparison or not, results in immediate program termination\cite[\S3.1]{Haskell98}.}. Therefore, the return value would likely be \textsf{Some None}, rather than \textsf{Some null}.  Since \textsf{null} is not used, there is never any danger of a null value being dereferenced.  Functional languages plan for failure explicitly; imperative languages imply failure implicitly.

Note that I have used a typically functional feature known as \emph{pattern matching} to evaluate a result based on the \emph{structure} of the result, rather than the content of the result.  It does not matter what the value that is ``wrapped'' within the \textsf{Some} case is.  All that matters is that, structurally, a \textsf{Some} case is distinguished from a \textsf{None} case.  This emphasis on the structure of data, rather than the content of data (such as the current value of a variable), is a line of thought that will be seen in other features as well.

\subsection*{Fail to a Secure Mode}

In \cite[p.~65]{WritingSecureCode}, Howard \& LeBlanc provide the following pseudocode to describe a failure of this principle:

\begin{verbatim}
DWORD dwRet = IsAccessAllowed(...);
if (dwRet == ERROR_ACCESS_DENIED) {
   // Security check failed.
   // Inform user that access is denied.
} else {
   // Security check OK.
   // Perform task.
}
\end{verbatim}

They challenge the reader to find the problem with the above code, which seems to be quite secure at first glance.  In fact, the issue is in the comparison \textsf{dwRet~==~ERROR\_ACCESS\_DENIED}.  Secure code should only execute when \textsf{dwRet} is equal to \textsf{NO\_ERROR}; in the imperative code presented, any return code other than \textsf{ERROR\_ACCESS\_DENIED} will result in the secure path being taken.  This approach of checking against known-good values and denying all other values is called \emph{whitelisting} or \emph{deny-by-default} in other spheres of computer science. Imperative languages don't make whitelisting easy, however; in fact, by making it easy to return from a function at any point, they encourage the practise of checking for certain \emph{invalid} values early on, and returning an error if any of those values is detected.

The same code could be expressed much more safely in a functional manner.  Given the following discriminated union:

\begin{verbatim}
type AccessLevel =
| Administrator of CredentialInfo
| PowerUser of int * CredentialInfo
| NormalUser of CredentialInfo
| Unauthorised
\end{verbatim}

...the following code is easy to write:

\begin{verbatim}
let level = levelOf user
match level with
| Administrator _ -> (* secure code here *)
| PowerUser (n,_) when n > 9000 -> (* alternative secure path *)
| _ -> (* reject request *)
\end{verbatim}

The most important feature of this code is that the secure paths --- \textsf{Administrator}-level access, and \textsf{PowerUser} access when the level is above 9000 --- are whitelisted.  All other paths are rejected by the default case.  The above code is also the easiest and most idiomatic functional code to write, which means that whitelisting is the default position in a functional language.  By contrast, code which does not whitelist is more difficult to write:

\begin{verbatim}
let level = levelOf user
match level with
| NormalUser _ | Unauthorised -> (* reject *)
| Administrator _ -> (* secure code here *)
| PowerUser (n,_) when n > 9000 -> (* alternative secure path *)
| PowerUser _ -> (* reject *)
\end{verbatim}

Not only is this more to write, and includes duplicate rejection code, the language compiler will generally complain if the \textsf{AccessLevel} discriminated union is changed and the non-whitelisting code is not updated.  A programmer quickly feels the maintenance burden and duplication burden of insecure code, and is therefore encouraged to take the secure path by default.

\subsection*{Assume External Systems Are Insecure}

The same kind of whitelisting mechanism that is used to fail securely can be used to parse, check, and potentially fail-securely with regard to incoming data from an external system.  There are two types of input that are considered in this section: structured input, such as may be obtained from a third-party web service, and unstructured input, such as console or text-box input.

The mechanism itself is a simple extension of the pattern matching concept: the input is parsed by a function into a structured type, which is then pattern matched against.  In F\#, this mechanism is called \emph{active patterns}\cite{ActivePatterns,MSDNActivePatterns}; in Haskell, it is implemented as a compiler extension to the reference compiler and called \emph{view patterns}\cite{HaskellViewPatterns}; in Scala, the feature is called \emph{extractor patterns}\cite{ProgrammingInScala}; and in OCaml, the macro system provides a superset of the required functionality.

Converting structured input into a pattern-matchable form is a problem that arises only in languages that interoperate or integrate closely with with non-functional code; languages that don't interoperate easily are presented with the problem of parsing unstructured input (such as bytes from the network) into a structured form instead.  To handle structured input, it is sufficient to transform the input into a form that is suitable for pattern-matching.

\begin{verbatim}
type StructuredRecord = { Name : string; Age : int }

let (|Structured|_|) (x : ConventionalObject) =
   if x = null then None
   else Some { Name=x.Name; Age=x.Age }

let greet o =
   match o with
   | Structured { Name="Bruce" } -> "Welcome back, Mr Wayne."
   | Structured { Age=a } when a < 25 -> "Wassup?"
   | Structured { Name=n } -> sprintf "Hello %s!" n
   | _ -> "Couldn't convert object; failure case here."
\end{verbatim}

In the above example, a conventional object (of type \textsf{ConventionalObject}) is converted into a \emph{record type} by the active pattern\footnote{Strictly speaking, the example shows a \emph{partial} active pattern; even simpler active patterns with discriminated cases are possible (see \cite{MSDNActivePatterns})} \textsf{Structured}.  If the object cannot be converted -- in this case, if it is \textsf{null} -- the pattern match falls through to the default case, where failure would be handled securely.  The active pattern could also have converted \textsf{ConventionalObject} to a tuple, discriminated union, or any other pattern-matchable object.

For unstructured input, we can parse securely using a variation\footnote{Based on a code sample presented in \cite[p.~32--3]{ActivePatterns}.} on the above.

\begin{verbatim}
let (|Match|_|) pattern input =
   let m = Regex.Match (input, pattern)
   if m.Success then Some (List.tail [for x in m.Groups -> x.Value])
   else None

let identify input =
   match input with
   | Match @"^Username=(\w+)$" [user] -> printfn "User is %s" user
   | Match @"^Name=(\w+)\s+(\w+)$" [name;surname] -> printfn "Full name is %s %s" name surname
   | _ -> (* handle failure *)
\end{verbatim}

Once again, we specify the whitelisted cases, extract the data which matches the whitelist, and any input that does not match is handled by the default failure case.

\subsection*{Minimise Your Attack Surface}

The ``attack surface'' of an application is defined by example in \cite[p.~57]{WritingSecureCode} as including sockets, named pipes, RPC\footnote{\textbf{R}emote \textbf{P}rocedure \textbf{C}all} endpoints, services, dynamic web pages, accounts, files, directories, and registry keys --- as well as any other possible vectors.  Manadhata lays out a much more sophisticated and systematic analysis of what the term ``attack surface'' means for a programmer in his 2008 thesis\cite[\S3.2.7]{AttackSurfaceMetric}:

\begin{quote}
Given a system, $s$, and its environment, $E_s$, $s$'s attack surface is the triple $\langle{}M^{E_s}, C^{E_s}, I^{E_s}\rangle{}$, where $M^{E_s}$ is the set of entry and exit points, $C^{E_s}$ is the set of channels, and $I^{E_s}$ is the set of untrusted data items of $s$.
\end{quote}

Entry points may be direct, which means that they accept input directly from the environment, or indirect, which means that they accept input from a chain of inputs which terminates in the environment.  Exit points may similarly be direct or indirect.  A channel is a path by which entry and exit points receive and send data respectively, such as a TCP socket or web request/response.  Lastly, an untrusted data item is ``a persistent data item $d$ such that a direct entry point of $s$ reads $d$ from the data store or a direct exit point of $s$ writes $d$ to the data store''\cite[p.~26]{AttackSurfaceMetric}, where a \emph{persistent} item is visible to both system and user across different executions of $s$.

The most obvious entry and exit points are found in the public Application Programming Interface (API) of a software component.  In this regard, there appear to be no significant differences between the size of a functional API and the size of a non-functional API.  For example, the C\# \textsf{System.Collections.Generic.Dictionary}\cite{MSDNDictionary} type API consists of 7 constructors, 9 public methods\footnote{Excluding 5 public methods inherited from \textsf{System.Object} but not overloaded.}, and 5 properties for a total of 21 entry/exit points; by contrast, the \textsf{Microsoft.FSharp.Collections.Map}\cite{MSDNMap} module contains 25 functions, which are its entry/exit points.  There appears to be no attack surface advantage that accrues to functional programming.

The situation changes once multi-core programming is considered.  Recall the triple $\langle{}M^{E_s}, C^{E_s}, I^{E_s}\rangle{}$: in a thread-using imperative program, $C^{E_s}$ includes mutable thread-accessible memory, $M^{E_s}$ includes all reads and writes to thread-shared variables, and $I^{E_s}$ includes all variables that threads can potentially alter.  These variables need to be synchronised and protected to avoid deadlocks, livelocks, and race conditions.

\begin{verbatim}
void Expirer() {
    List<K> to_expire = new List<K>();
    while (true) {
        Thread.Sleep(expiry);
        to_expire.Clear();
        lock (storage) {
            foreach (var kvp in storage) {
                if (DateTime.Now - kvp.Value.Item2 >= expiry)
                    to_expire.Add(kvp.Key);
            }
        }
        foreach (var k in to_expire)
            lock (storage) { storage.Remove(k); }
    }
}
\end{verbatim}

The code sample above could be part of a very basic C\# timed-expiry cache.  It shows the thread within the cache that expires items at the appropriate time.  Note that the \textsf{storage} dictionary must be held exclusively while the thread does cache expiry.  As a performance optimisation, the programmer releases \textsf{storage} between removals of each expired item.  This has the effect of allowing an attacker to insert a key at the appropriate timing and have it almost instantly removed by the system; whether this unintended consequence has any effect on security depends on the system that the flawed cache is used in.  However, what is important to note is that an entry point into the system can allow a value to be removed from an untrusted data store: the attack surface of the multi-threaded program is increased.

\begin{verbatim}
let cache timeout =
   let expiry = TimeSpan.FromMilliseconds (float timeout)
   let exp = Map.filter (fun _ (_,dt) -> DateTime.Now-dt >= expiry)
   let newValue k v = Map.add k (v, DateTime.Now)
   MailboxProcessor.Start(fun inbox ->
      let rec loop map =
         async {
            let! msg = inbox.TryReceive timeout
            match msg with
            | Some (Get (key, channel)) ->
               match map |> Map.tryFind key with
               | Some (v,dt) when DateTime.Now-dt < expiry ->
                  channel.Reply (Some v)
                  return! loop map
               | _ ->
                  channel.Reply None
                  return! loop (Map.remove key map)
            | Some (Set (key, value)) -> return! loop (newValue key value map)
            | None -> return! loop (exp map)
         }
      loop Map.empty
   )
\end{verbatim}

Though the functional world has not yet settled on a dominant paradigm for multi-core programming -- for example, Haskell prefers software transactional memory\cite{HaskellSTM}, and OCaml researchers are still investigating alternatives -- one of the oldest patterns for multi-core programming is found in Erlang\cite{Erlang}, and implemented in Scala\cite{ScalaActors} and F\#: the actor model.  In this pattern, messages are sent to ``actors'', who take some action based on the message content and, if requested, return some reply to the sender.  Messages and responses are immutable, and the state within an actor is kept in an immutable data structure; it is impossible for the flaw in the imperative cache system to exist in the functional implementation.  Immutability means that the attack surface for a single-core program is the same as the attack surface for a multi-core program; the same cannot be said for the imperative paradigm.

\subsection*{Employ Secure Defaults / Backward Compatibility Will Always Give You Grief}

This principle refers to removing uncommonly-used or insecure-by-default features from a default installation, or deprecating/removing those features altogether.  A ``feature'' is some functionality that makes up part of a system; in the imperative object-oriented world, creating a new feature may involve creating new classes, determining the interactions between those classes, and integrating the classes with existing code.  At a minimum, a new feature involves the alteration of existing methods to provide different functionality.

Given this common-sense description of what a feature is, it follows that the granularity of a system --- the extent to which features can be decomposed and regarded as separate units of functionality --- depends largely on the coupling of features to the surrounding code of the system.  Features that are tightly-coupled to the surrounding code are difficult to disable since removing them requires a large number of changes to the surrounding code.  Conversely, features that are loosely-coupled require only a small amount of work to disable.

Different methods have been proposed for measuring coupling.  Green et al.\cite[p.~23--9]{SliceIntroduction} provide a good summary of the past 25 years of work on slice-based methods\footnote{These methods follow on from, and may be considered to be superior to, the metrics proposed by Henry \& Kafura in \cite{HenryAndKafura}; see, for example, the discussion of flow \emph{vs} bandwidth in \cite[p.~8]{EmpiricalSlicingStudy} or \cite[p.~23]{SliceIntroduction}.} for determining program coupling.  Briefly, a \emph{backward slice} is the set of program statements that affect the calculation of a variable, and a \emph{forward slice} is the set of program statements that are affected by the calculation of a variable.  In layman's terms, to calculate the coupling for a module one starts at \emph{output variable}s --- which may be a function return values, global variables, printed variables, or reference parameters --- and determines the full backward slice for that variable, which provides all the calculations that it is dependent on.  It is then easy to see which dependent variables are found in other modules and to determine the extent of the coupling.

While slicing can be generalisable and applicable to object-orientated languages, its roots are found in traditional imperative languages such as C.  Chidamber \& Kemerer proposed the \emph{Coupling Between Objects} (CBO) metric\cite[p.~485--7]{OOMetricsSeminal}, defined as usage of methods or attributes of a class $A$ from a class $B$.  Briand et al. provide an overview of coupling metrics in \cite{UnifiedCoupling}, some of which are more complex than Chidamber \& Kemerer's understanding, and some of which are simpler.  All coupling metrics agree on the CBO metric as a baseline, and differ on which other aspects of the object-oriented approach (such as inheritance) affect coupling.

It is clear, from the body of work on coupling, that this measure of complexity is considered important, with ``more than 30 different measures of object-oriented coupling''\cite[p.~3]{UnifiedCoupling} found extant in 1999.  It is taken as a given that coupling can be mitigated, but not eliminated in non-trivial projects.  Many authors have stated that tightly-coupled modules are difficult to untangle --- and, consequently, tightly-coupled features are difficult to disable.  Furthermore, attempting to disable or remove a tightly-coupled feature is an error-prone task.

The fundamental reason for the existence of coupling measures is implied by slice-based analysis: the value of an output variable depends on the calculation of dependent variables, and affects the calculation of other variables in turn.  This reveals, once again, the emphasis that the imperative paradigm places on the importance of variable \emph{value}, as opposed to data \emph{structure}.  In object-oriented terms, the problems that coupling points to are a microcosm of this effect:

\begin{itemize}
\item two invocations of the same method on a class may return entirely different results, depending on the current state of the class
\item the state of a class may be altered by other functions (in C\# and Java, this is accomplished via ``getters'' and ``setters'' in programs that follow best practice)
\end{itemize}

Coupling is not an active research area in the functional programming space.  As a consequence of immutability, pure functions always return the same value when provided with the same input parameters; there is no ``local state'' that can affect the calculation.  This is known as \emph{referential transparency}\label{term:ReferentialTransparency}.  Actors are an exception to this norm: when using the actor model for multi-core programming, it can be argued that an actor could contain some local state.  This state is immutable: more specifically, it can be said that there is no state that changes, but there is a different immutable state that exists for each request and which cannot be affected by anything other than the actor.  Inheritance, global mutable variables, mutable reference parameters, and other aspects of the imperative or object-oriented world that lead to coupling are largely unknown.

The immutability of data and its consequence for pure functions means that features are, by default, loosely-coupled in a functional system.  Disabling or removing features to achieve secure defaults is correspondingly easier.

\subsection*{Fix Security Issues Correctly / Learn From Mistakes}

To fix security issues correctly or learn from mistakes, a programmer should be able to understand the reasoning behind the program statements. Consider the following imperative code:

\begin{verbatim}
public class LargeItem {
	...
    public Entity Mangle(Entity e, int n) {
        if (!Reevaluate(e)) return null;
        if (n > 0x33 && n < 0x55 && n != 0x44)
            e.Discombobulate(0x55, n, 0x33);
        return e;
    }
    ...
}
\end{verbatim}

A typical call to this code could look something like

\begin{verbatim}
large.Mangle(someEntity, input).Zombify();
\end{verbatim}

What could go wrong with this code?

\begin{itemize}
\item The call to \textsf{Mangle} could return \textsf{null}, causing the call to the \textsf{Zombify} member to throw a null-reference exception.

\item \textsf{Reevaluate} could alter the value of the passed \textsf{Entity}, which could change the result of the \textsf{Discombobulate} call.

\item \textsf{Mangle} will always perform a call to \textsf{Reevaluate}, but will only sometimes perform a call to \textsf{Discombobulate}.  If the caller expects \textsf{Discombobulate} to be called for every invocation, this could lead to unexpected behaviour.

\item \textsf{Reevaluate} will succeed or fail based on the internal state of the \textsf{LargeItem} instance, which may or may not be appropriate at the time of the call.

\item The range-check for the first and third parameters of \textsf{Discombobulate} are performed in \textsf{Mangle}, outside of the body of \textsf{Discombobulate}.
\end{itemize}

This (incomplete) list should be sufficient to show some of the issues that could arise when determining why a problem has arisen at the call-site.  Diagnosing any of the behaviours listed requires analysis of all the values that could affect the computation at the call-site, and one of the easiest ways to do this is by attaching a debugger and examining values.

Given industry demands for productivity (and, quite possibly, programmer ego!) there is an incentive to ``resolve'' bugs as quickly as possible.  For example, if the issue is a null-reference exception, the easiest fix is to write an \textsf{if}-statement to check for the \textsf{null} value, and only call the \textsf{Zombify} member if there isn't a \textsf{null} return.  The further down the call-tree the programmer goes, the more dependent variables there are to consider, and the more difficult the ``right'' solution is to determine --- let alone fix correctly.  Reasoning about the code becomes increasingly difficult and dependent upon runtime values.

The corresponding functional code may look as follows:

\begin{verbatim}
let mangle n relatedInfo entity =
   entity
   |> reevaluate relatedInfo
   |> Option.map (discombobulate n)
...
let mangled =
   someEntity
   |> mangle input related
mangled |> Option.map zombify
\end{verbatim}

A few points are worth mentioning about this code, as compared to the imperative version of the same logic:

\begin{itemize}
\item The state information by which the imperative \textsf{Mangle} would determine the result of \textsf{Reevaluate} is explicitly encoded as a parameter in the functional version.  Therefore, all items that could affect the calculation of \textsf{mangled} can be identified by visual inspection.

\item The value of \textsf{someEntity} remains the same throughout.  The transformed value may be used explicitly by referencing the \textsf{mangled} symbol instead of the \textsf{someEntity} symbol.

\item No null-reference exception is possible (see ``Plan On Failure'', p.~\pageref{sec:PlanOnFailure}).

\item \textsf{discombobulate} cannot have any incidental side-effects (such as resetting a boolean flag) since mutable data does not exist in the functional paradigm.  Whether it is called or not is only relevant in terms of the transformation of \textsf{someEntity}.

\item The range-check has been shifted into the body of \textsf{discombobulate} (not shown).  If it were \emph{not} shifted, the ``pipelining'' syntax of \textsf{mangle} would be interrupted, and the code would be less idiomatic; I believe that pipelining syntax visually encourages the programmer to shift conditionals towards their correct place.
\end{itemize}

The most important difference, however, is that it is possible for a programmer to reason about the program that is being written \emph{without} reference to runtime variable values, and with reference to the structure of the data that is being transformed.  This makes it easier to determine why a particular issue has arrived, pinpoint the cause of the issue, and apply the correct fix.

\subsection*{Use Defense in Depth / Never Depend on Security Through Obscurity Alone}

Defense in depth involves layering ``defensive'' code (such as authorisation checks, resource-scoping code, and input/output-encoding) around the system, with the goal of stopping an attacker who has circumvented $n$ layers at layer $n+1$.  Some features of the functional paradigm lend themselves to this easily: for example, the fact that data is immutable makes it extraordinarily difficult for an attacker to modify a ``shared'' resource (such as a path to ``authorised'' plugin objects), since there are no mutable shared resources.

However, it is one thing to have features that enhance security, and quite another to arrange for them to be layered as defenses that must be broken through by a determined attacker.  In an imperative object-oriented language, one would trade readability for security with a construction such as

\begin{verbatim}
StaticMethods.EncodeOutput(
   StaticMethods.CheckCredentials(
      StaticMethods.CheckAuthorisation(
         StaticMethods.ValidateToken(
            ...actual method call here...
         ))));
\end{verbatim}

In C\#, extension methods ease the syntax by removing the necessity of referencing \textsf{StaticMethods} at every turn; in Java, the \textsf{import static} directive has the same effect.  Nevertheless, the code is cumbersome to write.  At a minimum, all of the defensive measures might be placed into a separate static method, and the resulting defensive code might be as simple as

\begin{verbatim}
DefendOutput(...actual method call here...);
\end{verbatim}

If a programmer fails to call \textsf{DefendOutput}, no checks are done.  The alternative approach suggests placing the checks within the relevant methods, so that they done even if the caller doesn't call a separate method such as \textsf{DefendOutput}.

\begin{verbatim}
if (!StaticMethods.CheckCredentials(...)) return null;
if (!StaticMethods.CheckAuthorisation(...)) return null;
if (!StaticMethods.ValidateToken(...)) return null;
...
return StaticMethods.EncodeOutput(output);
\end{verbatim}

This transfers the burden of ensuring that security checks are done from client code to library code.  Recognising this, various frameworks for imperative languages, especially in the domain of web applications, have created special syntax and attributes to make the process of applying security checks easier.

Using a functional language, the situation becomes somewhat easier since functions can be easily created via \emph{composition}, \emph{chaining}, and \emph{partial application}.

Composition and chaining refers to the creation of a new function by combining existing functions.  One of the applications of this is creating ``chokepoints''\cite[p.~345--7]{WritingSecureCode}, which are functions that encode input or output.

\begin{verbatim}
let encodeInput (s : string) = (* take a string as input, return a string as output *)
let someFunction x = (* manipulate x and output a string *)
let someOtherFunction x y z = (* manipulate x, y, and z, and output a string *)
let encodedFunction = someFunction >> encodeInput
let encodedOtherFunction a b c = someOtherFunction a b c |> encodeInput
\end{verbatim}

In this example, \textsf{encodedFunction} is the result of composing \textsf{someFunction} and \textsf{encodeInput}; any input to \textsf{f } will be processed by \textsf{someFunction}, then by \textsf{encodeInput}, transparently to the caller.  \textsf{encodedOtherFunction} chains \textsf{someOtherFunction} to \textsf{encodeInput} --- composition is not possible since \textsf{someOtherFunction} and \textsf{encodeInput} take different numbers of parameters --- and is also transparent to the user.  Exposing only the composed/chained functions will ensure that all input and output is always encoded.

Partial application is a way to specify some of a function's arguments in order to specialise that function.

\begin{verbatim}
let isAuthorised user contextInfo functionality = true
let canDo user contextInfo = isAuthorised user contextInfo

// usage:
// canDo ViewReceipts
\end{verbatim}

\textsf{canDo} is a specialisation of \textsf{isAuthorised} for a specific user and context.  At the beginning of every invocation of functionality (for example, at the start of each web request), it is useful to create partially-applied functions which ``capture'' the current context, and pass these functions onwards for use elsewhere.  The resulting functions are more general in nature by virtue of being specialised: a caller of \textsf{canDo} does not need to know that \textsf{user} and \textsf{contextInfo} exist, and only needs to know which functionality (such as \textsf{ViewReceipts}) is required.

While there appears to be no good reason for chaining and composing functions to not be used as part of the imperative paradigm, the fact is that small functions which merely combine or specialise other functions are not idiomatic in most imperative languages.  I theorise that this is because some commonly-found features of functional languages are not found in mainstream imperative languages.

The first of these features is \emph{referential transparency}, which has already been discussed briefly on p.~\pageref{term:ReferentialTransparency}.  An implication of referential transparency is that any symbol can be replaced by the calculations used to obtain that symbol, with no change in the meaning of the program.  Since variables in the imperative paradigm are mutable, we cannot be sure that a symbol will remain equivalent to the result of its initial assignment; nor can we be sure that the variables which were used to calculate the symbol will not change between the calculation of the symbol and our use thereof.  Without this certainty, partial application of a function requires explicitly capturing the \emph{current} value of a variable, which (in turn) may entail creating a custom data structure -- quite a lot of trouble to take for some eventual notational convenience!  Chaining and composing functions causes some difficulty as well, since a call such as \textsf{DoX(DoY(data))} makes debugging more difficult: if an error occurs, is it as a result of \textsf{DoX} or \textsf{DoY}?  A programmer examining the issue would have to follow a variable into the \textsf{DoY} call, then into the \textsf{DoX} call, which breaks the line-based debugging that is a feature of many development environments.  Composing more functions, \textsf{DoX(DoY(DoZ(data)))}, exacerbates this issue.

The second of these features is the granularity of units of functionality.  In a functional language, the standard unit of self-contained functionality is the function.  It is dependent on nothing except for its input parameters, and may be called freely from any place that is able to reference it; there is no ``container'' that must be created before it can be used.  In an imperative non-object-oriented language, the standard unit of functionality is the program: all functions can modify program state, via references or global variables, and it is telling that one of the criticisms of non-object-oriented imperative languages is that they can easily lead to ``spaghetti code'' if not structured correctly\cite{SpaghettiCode}.  In object-oriented imperative code, the standard unit is the object: to do anything useful, the programmer has to create an object, and call methods on it.  To compose or chain methods in different non-static classes, a programmer must pass along references to each class; to simulate partial application, a design pattern such as Command\cite[p.~233--41]{DesignPatterns} may need to be used.  Due to the granularity of units of functionality being the class, rather than the function, combining functions is an unwieldy proposition.

Lastly, functional programming languages typically use a form of type inference based on the Hindley-Milner algorithm\cite{Hindley,Milner}, which allows for functions that support \emph{parametric polymorphism}.  This concept is best explained by an example:

\begin{verbatim}
let double x = x * 2
let result = List.map double [0;1;2;3;4]
\end{verbatim}

\textsf{result} will be the list \textsf{[0;2;4;6;8]}; the function \textsf{double} has been applied to each element of the input list \textsf{[0;1;2;3;4]} by the \textsf{List.map} function.  This seems reasonable, until we consider that \textsf{List.map} doesn't simply work on lists of integers; it can work on lists of any type.  Creating a function that works with any types is something that has been difficult in statically-typed imperative languages: Java has used boxing and type-erasure with compiler-time checks to solve it, C upcasted to \textsf{void*} and used the \textsf{sizeof} operator, C++ has developed the concept of templates, and C\# chose reified generics as their solution.  F\#, Haskell, Scala, and OCaml use Hindley-Milner type inference instead.  The type inference algorithm selects the most \emph{general} possible type for a symbol, given the way in which the symbol is used.  In the case of \textsf{List.map}, the function signature is

\begin{center}
\textsf{('a $\rightarrow$ 'b) $\rightarrow$ 'a list $\rightarrow$ 'b list}
\end{center}

In other words, given a function which takes a \textsf{'a} and returns a \textsf{'b}, and a list of \textsf{'a}, a list of \textsf{'b} will be returned; each of these parameters is generalised (or \emph{polymorphic}) to the maximum extent possible.  This process of generalisation occurs automatically.

By contrast, in most statically-typed object-oriented imperative languages, type inference does not play a large role.  Generalisation is explicit, along the lines of class hierarchies and interfaces: to generalise a method, one either passes in a superclass in the inheritance hierarchy, or extracts it into a named interface and then uses that interface as a parameter to the appropriate method.  Explicit generalisation makes combining methods into a manual exercise of matching types, which means that imperative programmers may see it as a needless and laborious task.

\subsection*{Don't Mix Code and Data}

The desire to mix code and data is one that arises when user control of a system, or user interactivity, is desired.  Lotus 1-2-3 version 2.0 is cited as one of the first widely-used systems to incorporate interactivity in the form of macros --- data which was interpreted as code --- and thus ``perform custom actions defined by the user''\cite[p.~67]{WritingSecureCode}.  A safer method of custom code execution is sandboxing, with many advances made in that area in recent times\cite{NaCl}.

Nevertheless, the desire to create some form of user interactivity that is less powerful (or more restricted) than native code execution, but more powerful than pre-built parameterised commands, may exist.  For example, a game developer may wish to allow users to script certain behaviours within a game, but not have any significant level of access to the API of the game.  This can be achieved by creating a domain-specific language (DSL), which is interpreted and gives rise to the desired results.

A DSL may be categorised as either internal or external\cite{FowlerDSLs} depending on whether the DSL is expressed in terms of the host language or not.  Internal DSLs are difficult to compare in any meaningful way since they necessitate a comparison of host language syntaxes and the manner in which they relate to the DSL that is being examined; such a comparison is fraught with subjectivity, since one programmer's verbosity is another programmer's explictness.  External DSLs don't suffer from this issue, and are more powerful cognitive tools in any case since they do not force a user to think in the syntax of a language that has not been built to match the problem domain.

In both imperative and functional languages, a parser for the DSL would be generated based on a formal grammar.  The parse results in an abstract syntax tree, and here the functional and imperative worlds diverge.  In an imperative language, the nodes of the tree must be examined by manual inspection or by creating classes for each important construct in the tree.  In a functional language, the tree is typically represented as a discriminated union, and pattern-matching is used to decompose and process it; the usual security benefits of whitelisting apply.

\subsection*{Use Least Privilege}

It is axiomatic that security should be embedded at the lowest possible level.  For example, if an operating system allows a programmer to lock a security-relevant resource for exclusive use, it makes no sense to create application-level code to repeatedly test whether the resource has been modified; another application may circumvent such code, or merely use the operating system functionality to go around it.  This does not contradict the idea of defense in depth, since in-depth security measures should be orthogonal.

The same low-level security measures that are available to imperative code are also available to functional code: access-control lists, permissions, locks, exclusive channels, and so on.  Some of these may not be used as much by idiomatic functional code --- for example, the actor model allows for lock-free multi-core programming, which makes mutexes irrelevant --- but whether a given measure is typically used or not has no impact on its availability for use.  This is as true for privilege-affecting security measures as it is for more general security measures.

Implementing the principle of least privilege in an imperative language involves data flow analysis as a first step\cite[p.~61, 73--5]{WritingSecureCode}.  \emph{A priori}, we can understand the value of data-flow analysis as soon as we accept that different resources should have different levels of access associated with them.  For example, anybody may be able to read a network-shared public file; creating a trust relationship between two networked machines should intuitively be a more privileged operation.  Furthermore, the type and context of access matters: reading a file from a desktop application and writing a file from a web browser are two very different scenarios.

Schwartz et al. point out the difficulties inherent in conventional data flow analysis by formalising existing literature around dynamic\footnote{Static taint analysis involves no runtime checks, and is not considered under in this paper.} taint analysis\cite{DynamicTaintAnalysis}.  Taint analysis revolves around the idea that input from an untrusted source is ``tainted'', and should not be used until it has been validated in some way.  Howard \& LeBlanc devote an entire chapter to this idea\cite[p.~341--62]{WritingSecureCode}, underscoring its importance.  This does not mean that data flow analysis is impossible: it is an area of active research, and some of the tools that have been developed for dynamic taint analysis are promising\cite{DataFlowAnalysis,DataFlowAssertions}.

In a functional language, the data flow is easier to perceive, and programmers rarely need tools to explicate it.  This is partly because of a feature that has already been discussed: discriminated unions.  Most dynamic taint analysis tools work by tagging security-relevant data\cite{DynamicTaintAnalysis,DataFlowAnalysis} and examining them at crucial points; there is no need for an additional tool to do this when the ability to tag and differentiate data is built into the language.

Discriminated unions do not account for the majority of the data flow clarity of functional languages, however.  \emph{Higher-order} functions, of which we have already encountered one (\textsf{List.map}), accounts for the rest.  A higher-order function is simply a function which uses a function as input or output.  Using higher-order functions allows a programmer to perform common operations with the least code possible.  For example, it is frequently the case that a single value must be calculated on the basis of a sequence of values: any imperative function which returns a scalar value and contains a loop would fall into this category.  In C\#, such a function might look like

\begin{verbatim}
public double AveragePrice(double[] competitors) {
    double average = 0.0;
    for (int i = 0; i < competitors.Length; ++i) {
        average = ((average * i) + competitors[i]) / (i + 1);
    }
    return average;
}
\end{verbatim}

The corresponding F\# could be

\begin{verbatim}
List.fold (fun (i,average) price ->
   i+1., (average*i + price) / (i+1.)
) (0.,0.) >> snd
\end{verbatim}

A few points arise as the two versions are compared.

\begin{itemize}
\item The imperative version gives little indication that it operates on a sequence and uses each item in that sequence to arrive at a particular value.  It \emph{could} be implied by the method name, the scalar return value, and the array of values passed as a parameter, but the method could also take into account class state or some other data.  Without looking at the body of the method, it is impossible to know for sure.  On the functional side, the mere fact that a function called \textsf{fold} has been invoked tells the reader immediately that a sequence of values is being reduced to a single value.  Other common names in functional programming, such as \textsf{map} or \textsf{filter}, provide similar recognizability of the operation being applied: the data flow can be understood instantly by noticing the name of the higher-order function.

\item The core of imperative logic --- the average calculation itself --- is passed through as a function in the functional version, and is visually next to the invocation of \textsf{fold}.  A reader wishing to know \emph{how} the reduction to a single value occurs has the answer immediately available.  In the imperative version, this is located in the middle of the loop, and is not immediately obvious.

\item The initial value to use during the calculation is declared at the top of the imperative version, though the reader has no idea of its importance until the \textsf{return} statement is seen or the calculation itself is examined.  In the functional version, the initial value \textsf{(0.,0.)} is provided next to the core logic.

\item There is no explicit indexing in the functional version.  In the imperative version, the \textsf{i} variable performs a dual task: indexing the \textsf{competitors} array, and counting the number of items that have been seen thus far.  Given a different loop algorithm --- for example, one that requires the variable to start from 1 instead of 0 --- it is easy for a programmer to make either a calculation mistake or an indexing mistake, both of which could have security implications.  Since there is only one purpose for the \textsf{i} symbol in the functional version, this mistake is more difficult to make.
\end{itemize}

The functional advantage in the case of least-privilege is a subtle one: it is easier to understand data flows in a functional language, and correspondingly easier to understand where to apply which level of privilege.

\subsection*{Remember That Security Features != Secure Features}

This principle states that security should be built in to features, rather than being ``tacked-on'' at the end of a development cycle.  On the program level, building secure features requires understanding of the flow of data and the reasons for that flow, rather than understanding each statement in isolation.  Alan Perlis referred to this in his 1982 epigram, ``A programming language is low level when its programs require attention to the irrelevant''\cite{PerlisEpigrams}.  As the examples from the previous section show, this understanding is easier to obtain when using a functional language; therefore, secure features are correspondingly easier to develop, and insecure features are easier to detect.

It can be said that building security in is a matter of choosing the secure option whenever a choice is presented.  In a functional language, the default loose coupling of the functional paradigm make dependencies explicit, so that any feature that is determined to be ``insecure'' can be removed or disabled with minimal trouble.  Choosing to use whitelists instead of blacklists is a security-conscious decision, and it is one that the functional paradigm makes easy.  Iterating through objects rather than indexing through a collection is a more security-conscious way of programming, as is explicitly indicating failure.  Importantly, an imperative language tends to regard these choices as having \emph{equal value}; a functional language values correctness (in the mathematical sense), and its constructs typically reflect a push towards that.

\section*{Caveats and possibilities}

Languages differ, whether they are functional or imperative.  Not all languages are statically-typed, for example: Erlang is a functional language that is dynamically-typed, just as Ruby and Python are dynamically-typed imperative languages.  Frameworks, libraries, syntax, and language philosophy (among other concerns) make any comparison between imperative languages difficult, and the same is true of functional languages.

Even the distinction between functional and imperative can become blurred: since both F\# and OCaml allow data to be mutable, are they really ``functional'' languages?  Javascript is considered to be an imperative language, but CoffeeScript\cite{CoffeeScript} (which compiles to Javascript) has functionally-inspired features; which is it?  To overcome this difficulty of categorisation, any languages which support immutability by default, higher-order functions, and referential transparency --- all of which are associated with the functional paradigm --- have been considered as functional languages, whether statically- or dynamically-typed.

I have compared statically-typed languages in this paper since I am most familiar with them.  Many of the points that have been made should be equally applicable to dynamically-typed languages, and most of the examples should be understandable and portable across language boundaries.  The aim of this paper has been to compare the functional \emph{paradigm}, as evidenced by features found in typical functional languages, with the imperative paradigm.  It has not been intended as a comprehensive language comparison between C\# and F\#, nor should it be mistaken for such.

\subsection*{A failure of vision?}

From the case presented, it seems clear that functional languages may have certain security advantages over imperative languages.  Empirical studies could validate this idea; as a reference point, the 2011 study \cite{ImperativeDefects} examines

\begin{quote}
...data from 362 projects of four different firms. These projects spanned a wide range of programming languages, application domain, process choices, and development sites spread over 15 countries and 5 continents.
\end{quote}

The mean defect density\footnote{Number of defects per thousand lines of code (defects/KLOC)} was $\frac{1}{10.06} \approx 0.1\pm0.04 $, with a maximum of $\frac{1}{0.01} \approx 100$ and minimum $\frac{1}{240.05} \approx 0.004$\cite[p.~265]{ImperativeDefects}.  Although a breakdown of defect density by language is not provided, the authors list Java, .Net, PHP, C, C++, and assembly language as the set of languages that were used\cite[p.~263]{ImperativeDefects}; all of these are imperative\footnote{Data was collected up to and including 2009.  F\# was only included as an official part of the .Net ecosystem in 2010.}.

No comparable study for the functional paradigm exists.  The most relevant comparison point appears to be \cite{HaskellDefectDensityMultiple}, which tracked defect density in the GHC Haskell compiler over a number of versions.  The density varied from 0.49 to 0.04, with a startling decline from 0.29 (GHC version 5.02.2) to 0.04 (GHC version 5.04).  Subsequent to version 5.04, the maximum defect density has been 0.08.  The authors of \cite{HaskellDefectDensityMultiple}, Sherriff et al., provide no explanation for this sharp drop, nor do they explain their methodology for determining what a ``defect'' is, other than saying that they used ``detailed documentation and defect logs'' \cite[p.~2]{HaskellDefectDensityMultiple}.  Furthermore, given that the chosen project was the reference Haskell compiler, it might be assumed that the programmers on the project are expert Haskell programmers: this may bias the defect density in the project's favour.

It is clear that more empirical research must be done in this area before any hard conclusions can be drawn.  That research is all the more difficult to do since functional languages are not as widely-used as imperative languages.  It is my hope that this paper will provide some incentive for security-conscious programmers and organisations to seriously investigate the use of the functional paradigm, and thus move the field towards a better understanding of it and its relationship to the existing imperative model.

\bibliographystyle{IEEEtran}
\bibliography{Functional_programming_and_security}

\begin{thebibliography}{10}
\providecommand{\url}[1]{#1}
\csname url@samestyle\endcsname
\providecommand{\newblock}{\relax}
\providecommand{\bibinfo}[2]{#2}
\providecommand{\BIBentrySTDinterwordspacing}{\spaceskip=0pt\relax}
\providecommand{\BIBentryALTinterwordstretchfactor}{4}
\providecommand{\BIBentryALTinterwordspacing}{\spaceskip=\fontdimen2\font plus
\BIBentryALTinterwordstretchfactor\fontdimen3\font minus
  \fontdimen4\font\relax}
\providecommand{\BIBforeignlanguage}[2]{{%
\expandafter\ifx\csname l@#1\endcsname\relax
\typeout{** WARNING: IEEEtran.bst: No hyphenation pattern has been}%
\typeout{** loaded for the language `#1'. Using the pattern for}%
\typeout{** the default language instead.}%
\else
\language=\csname l@#1\endcsname
\fi
#2}}
\providecommand{\BIBdecl}{\relax}
\BIBdecl

\bibitem{HistoryOfLambdaCalculus}
F.~Cardone, J.~Hindley, and N.~MRRS, ``History of lambda-calculus and
  combinatory logic,'' \emph{Handbook of the History of Logic}, vol.~5, 2006.

\bibitem{Tevis2004}
J.-E. Tevis and J.~Hamilton, ``Methods for the prevention, detection and
  removal of software security vulnerabilities,'' in \emph{Proceedings of the
  42nd annual Southeast regional conference}.\hskip 1em plus 0.5em minus
  0.4em\relax ACM, 2004, pp. 197--202.

\bibitem{SecureFuncParadigm}
J.-E. Tevis, ``Secure programming using a functional paradigm,'' in
  \emph{Proceedings of the Illinois State Academy of Science (ISAS)
  Conference}.\hskip 1em plus 0.5em minus 0.4em\relax Citeseer, 2006.

\bibitem{FuncPurityJava}
M.~Finifter, A.~Mettler, N.~Sastry, and D.~Wagner, ``Verifiable functional
  purity in {J}ava,'' in \emph{Proceedings of the 15th ACM conference on
  Computer and communications security}.\hskip 1em plus 0.5em minus 0.4em\relax
  ACM, 2008, pp. 161--174.

\bibitem{SecureProgrammingCookbook}
J.~Viega and M.~Messier, \emph{Secure Programming Cookbook for C and
  C++}.\hskip 1em plus 0.5em minus 0.4em\relax O'Reilly Media, 2003.

\bibitem{WritingSecureCode}
M.~Howard and D.~LeBlanc, \emph{Writing Secure Code}, 2nd~ed.\hskip 1em plus
  0.5em minus 0.4em\relax Microsoft Press, 2003.

\bibitem{OWASPSecureCodingPrinciples}
OWASP, ``{OWASP} {S}ecure {C}oding {P}rinciples,''
  \url{https://www.owasp.org/index.php/Secure_Coding_Principles}, 2009.

\bibitem{SDLArticle}
M.~Howard, ``How do they do it? a look inside the security development
  lifecycle at microsoft,'' \emph{MSDN Magazine}, pp. 107--114, 2005.

\bibitem{CSharpIsFunctional}
A.~Kennedy, ``C\# is a functional programming language,'' in \emph{Fun in
  Oxford Meeting}, 2006.

\bibitem{JavaPerformanceTuning}
J.~Shirazi, \emph{Java performance tuning}, 2nd~ed.\hskip 1em plus 0.5em minus
  0.4em\relax O'Reilly Media, 2003.

\bibitem{ExperimentingWithExceptions}
A.~Gorbenko, A.~Romanovsky, V.~Kharchenko, and A.~Mikhaylichenko,
  ``Experimenting with exception propagation mechanisms in service-oriented
  architecture,'' in \emph{Proceedings of the 4th international workshop on
  Exception handling}.\hskip 1em plus 0.5em minus 0.4em\relax ACM, 2008, pp.
  1--7.

\bibitem{EfficientExceptions}
R.~Orr, ``Efficient exceptions?'' \emph{Overload}, no.~61, pp. 15--20, 2004.

\bibitem{Haskell98}
S.~{Peyton-Jones}, \emph{Haskell 98 Language and Libraries: The Revised
  Report}.\hskip 1em plus 0.5em minus 0.4em\relax Cambridge University Press,
  2003.

\bibitem{ActivePatterns}
D.~Syme, G.~Neverov, and J.~Margetson, ``Extensible pattern matching via a
  lightweight language extension,'' in \emph{Proceedings of the 12th ACM
  SIGPLAN international conference on Functional programming}.\hskip 1em plus
  0.5em minus 0.4em\relax ACM, 2007, pp. 29--40.

\bibitem{MSDNActivePatterns}
{Microsoft Corporation}, ``Active {P}atterns ({F}\#),''
  \url{http://msdn.microsoft.com/en-us/library/dd233248.aspx}, 2011.

\bibitem{HaskellViewPatterns}
S.~{Peyton-Jones}, ``View{P}atterns -- {GHC},''
  \url{http://msdn.microsoft.com/en-us/library/dd233248.aspx}, 2011.

\bibitem{ProgrammingInScala}
M.~Odersky, L.~Spoon, and B.~Venners, \emph{Programming in Scala}.\hskip 1em
  plus 0.5em minus 0.4em\relax Artima Inc, 2008.

\bibitem{AttackSurfaceMetric}
P.~Manadhata, ``An attack surface metric,'' Ph.D. dissertation, University of
  North Carolina, 2008.

\bibitem{MSDNDictionary}
{Microsoft Corporation}, ``Dictionary {M}embers,''
  \url{http://msdn.microsoft.com/en-US/library/3eayzh46(v=VS.80).aspx}, 2011.

\bibitem{MSDNMap}
------, ``Collections.{M}ap {M}odule ({F}\#),''
  \url{http://msdn.microsoft.com/en-us/library/ee353880.aspx}, 2011.

\bibitem{HaskellSTM}
T.~Harris, S.~Marlow, S.~Jones, and M.~Herlihy, ``Composable memory
  transactions,'' \emph{Communications of the ACM}, vol.~51, no.~8, pp.
  91--100, 2008.

\bibitem{Erlang}
R.~Virding, C.~Wikstr{\"o}m, M.~Williams, and J.~Armstrong, ``Concurrent
  programming in erlang,'' 1996.

\bibitem{ScalaActors}
P.~Haller and M.~Odersky, ``Scala actors: Unifying thread-based and event-based
  programming,'' \emph{Theoretical Computer Science}, vol. 410, no. 2-3, pp.
  202--220, 2009.

\bibitem{SliceIntroduction}
P.~Green, P.~Lane, A.~Rainer, and S.~Scholz, ``An introduction to slice-based
  cohesion and coupling metrics,'' June 2009, technical Report No.488.

\bibitem{HenryAndKafura}
S.~Henry and D.~Kafura, ``Software structure metrics based on information
  flow,'' \emph{Software Engineering, IEEE Transactions on}, no.~5, pp.
  510--518, 1981.

\bibitem{EmpiricalSlicingStudy}
T.~Meyers and D.~Binkley, ``An empirical study of slice-based cohesion and
  coupling metrics,'' \emph{ACM Transactions on Software Engineering and
  Methodology (TOSEM)}, vol.~17, no.~1, pp. 1--27, 2007.

\bibitem{OOMetricsSeminal}
S.~Chidamber and C.~Kemerer, ``A metrics suite for object oriented design,''
  \emph{Software Engineering, IEEE Transactions on}, vol.~20, no.~6, pp.
  476--493, 1994.

\bibitem{UnifiedCoupling}
L.~Briand, J.~Daly, and J.~Wust, ``A unified framework for coupling measurement
  in object-oriented systems,'' \emph{Software Engineering, IEEE Transactions
  on}, vol.~25, no.~1, pp. 91--121, 1999.

\bibitem{SpaghettiCode}
G.~Cobb, ``A measurement of structure for unstructured programming languages,''
  \emph{ACM SIGSOFT Software Engineering Notes}, vol.~3, no.~5, pp. 140--147,
  1978.

\bibitem{DesignPatterns}
E.~Gamma, R.~Helm, R.~Johnson, and J.~Vlissides, \emph{Design patterns}.\hskip
  1em plus 0.5em minus 0.4em\relax Addison-Wesley, Reading, MA, 2002.

\bibitem{Hindley}
R.~Hindley, ``The principal type-scheme of an object in combinatory logic,''
  \emph{Transactions of the american mathematical society}, vol. 146, pp.
  29--60, 1969.

\bibitem{Milner}
R.~Milner, ``A theory of type polymorphism in programming,'' \emph{Journal of
  computer and system sciences}, vol.~17, no.~3, pp. 348--375, 1978.

\bibitem{NaCl}
B.~Yee, D.~Sehr, G.~Dardyk, J.~Chen, R.~Muth, T.~Ormandy, S.~Okasaka,
  N.~Narula, and N.~Fullagar, ``Native client: A sandbox for portable,
  untrusted x86 native code,'' \emph{Communications of the ACM}, vol.~53,
  no.~1, pp. 91--99, 2010.

\bibitem{FowlerDSLs}
M.~Fowler, ``A pedagogical framework for domain-specific languages,''
  \emph{Software, IEEE}, vol.~26, no.~4, pp. 13--14, 2009.

\bibitem{DynamicTaintAnalysis}
E.~Schwartz, T.~Avgerinos, and D.~Brumley, ``All you ever wanted to know about
  dynamic taint analysis and forward symbolic execution (but might have been
  afraid to ask),'' in \emph{2010 IEEE Symposium on Security and
  Privacy}.\hskip 1em plus 0.5em minus 0.4em\relax IEEE, 2010, pp. 317--331.

\bibitem{DataFlowAnalysis}
W.~Chang, B.~Streiff, and C.~Lin, ``Efficient and extensible security
  enforcement using dynamic data flow analysis,'' in \emph{Proceedings of the
  15th ACM conference on Computer and communications security}.\hskip 1em plus
  0.5em minus 0.4em\relax ACM, 2008, pp. 39--50.

\bibitem{DataFlowAssertions}
A.~Yip, X.~Wang, N.~Zeldovich, and M.~Kaashoek, ``Improving application
  security with data flow assertions,'' in \emph{Proceedings of the ACM SIGOPS
  22nd symposium on Operating systems principles}.\hskip 1em plus 0.5em minus
  0.4em\relax ACM, 2009, pp. 291--304.

\bibitem{PerlisEpigrams}
A.~Perlis, ``Special feature: Epigrams on programming,'' \emph{ACM SIGPLAN
  Notices}, vol.~17, no.~9, pp. 7--13, 1982.

\bibitem{CoffeeScript}
E.~Hoigaard, \emph{Smooth Coffee{S}cript}, 2011.

\bibitem{ImperativeDefects}
N.~Ramasubbu, M.~Cataldo, R.~Balan, and J.~Herbsleb, ``Configuring global
  software teams: a multi-company analysis of project productivity, quality,
  and profits,'' in \emph{Proceeding of the 33rd international conference on
  Software engineering}.\hskip 1em plus 0.5em minus 0.4em\relax ACM, 2011, pp.
  261--270.

\bibitem{HaskellDefectDensityMultiple}
M.~Sherriff, L.~Williams, and M.~Vouk, ``Using in-process metrics to predict
  defect density in haskell programs,'' in \emph{Fast Abstract, International
  Symposium on Software Reliability Engineering, St. Malo, France}.\hskip 1em
  plus 0.5em minus 0.4em\relax Citeseer, 2004.

\end{thebibliography}
\end{document}